\documentclass[a4paper,11pt]{article}
\usepackage{jcappub} 

\usepackage{lineno}
\usepackage{xcolor}
\usepackage{indentfirst}
\usepackage{physics}
\usepackage{comment}
\usepackage{amsmath}
\usepackage{mathtools}
\usepackage{siunitx}
\usepackage[position=b]{subcaption}

\newcommand{\I}{{$^{127}$I}}
\newcommand{\Na}{{$^{23}$Na}}

\newcommand{\CEvNS}{{CE$\nu$NS}}

\title{\boldmath Neutrino flux sensitivity to the next galactic core-collapse supernova in COSINUS}

\author[1]{G.~Angloher}
\author[1]{ M.~R.~Bharadwaj}
\author[2,3]{M.~Cababie}
\author[4,6]{I. Colantoni}
\author[5,6]{I.~Dafinei}
\author[5]{A.~L.~De~Santis$^*$}
\author[5,7]{N.~Di~Marco}
\author[2,3]{L.~Einfalt}
\author[7]{F.~Ferella}
\author[5,6]{F.~Ferroni}
\author[2]{S.~Fichtinger}
\author[7,8]{A.~Filipponi}
\author[1]{T.~Frank}
\author[2]{M.~Friedl}
\author[9]{Z.~Ge}
\author[10]{M.~Heikinheimo}
\author[1]{M.~N.~Hughes$^*$}
\author[10]{K.~Huitu}
\author[1]{M.~Kellermann}
\author[2,3]{R.~Maji}
\author[1]{M.~Mancuso}
\author[5,7]{L.~Pagnanini}
\author[1]{F.~Petricca}
\author[7]{S.~Pirro}
\author[1]{F.~Pr\"obst}
\author[7,8]{G.~Profeta}
\author[7]{A.~Puiu}
\author[2,3]{F.~Reindl}
\author[1]{K.~Sch\"affner}
\author[2,3]{J.~Schieck}
\author[2,3]{P.~Schreiner}
\author[2,3]{C.~Schwertner}
\author[1]{K.~Shera}
\author[1]{M.~Stahlberg}
\author[10]{A.~Stendahl}
\author[7,11]{M.~Stukel$^*$}
\author[7,12]{C.~Tresca}
\author[2,13,14]{F.~Wagner}
\author[9]{S.~Yue}
\author[1]{V.~Zema}
\author[9]{Y.~Zhu}
\author[7]{and G.~Pagliaroli}

\affiliation[1]{Max-Planck-Institut f\"ur Physik, 85748 Garching - Germany}
\affiliation[2]{Institut f\"ur Hochenergiephysik der \"Osterreichischen Akademie der Wissenschaften, 1050 Wien - Austria}
\affiliation[3]{Atominstitut, Technische Universit\"at Wien, 1020 Wien - Austria}
\affiliation[4]{Consiglio Nazionale delle Ricerche, Istituto di Nanotecnologia, 00185 Roma, Italy}
\affiliation[5]{Gran Sasso Science Institute, 67100 L' Aquila - Italy}
\affiliation[6]{INFN - Sezione di Roma, 00185 Roma - Italy}
\affiliation[7]{INFN - Laboratori Nazionali del Gran Sasso, 67100 Assergi - Italy}
\affiliation[8]{Dipartimento di Scienze Fisiche e Chimiche, Università degli Studi dell' Aquila, 67100 L' Aquila - Italy}
\affiliation[9]{SICCAS Shanghai Institute of Ceramics, Shanghai - P.R.China 201899}
\affiliation[10]{Helsinki Institute of Physics, 00014 University of Helsinki - Finland}
\affiliation[11]{SNOLAB, P3Y 1N2 Lively - Canada}
\affiliation[12]{CNR-SPIN c/o Dipartimento di Scienze Fisiche e Chimiche, Università degli Studi dell’Aquila, 67100 L’Aquila, Italy}
\affiliation[13]{Present address: Department of Physics, ETH Zurich, CH-8093 Zurich, Switzerland}
\affiliation[14]{Present address: ETH Zurich - PSI Quantum Computing Hub, Paul Scherrer Institute,CH-5232 Villigen, Switzerland}

\emailAdd{* alessio.desantis@gssi.it, hughes@mpp.mpg.de, matthew.stukel@snolab.ca}

\abstract{ 

While neutrinos are often treated as a background for many dark matter experiments, these particles offer a new avenue for physics:  the detection of core-collapse supernovae.
Supernovae are extremely energetic, violent and complex events that mark the death of massive stars. During their collapse stars emit a large number of neutrinos in a short burst. These neutrinos carry 99\% of the emitted energy which makes their detection fundamental in understanding supernovae.
This paper illustrates how COSINUS (Cryogenic Observatory for SIgnatures seen in Next-generation Underground Searches), a sodium iodide (NaI) based dark matter search, will be sensitive to the next galactic core-collapse supernova.
The experiment is composed of two separate detectors which will respond to far away and nearby supernovae. 
The inner core of the experiment will consist of NaI crystals operating as scintillating calorimeters. These crystals will mainly be sensitive to the Coherent Elastic Neutrino-Nucleus Scattering (CE$\nu$NS) against Na and I nuclei. The low mass of the cryogenic detectors enables the experiment to identify close supernovae within 1~kpc without pileup. The crystals will see up to hundreds of CE$\nu$NS events from a supernova happening at 200~pc. They reside at the center of a large cylindrical 230~T water tank, instrumented with 30 photomultiplier tubes. This tank acts simultaneously as a passive and active shield able to detect the Cherenkov radiation induced by impinging charged particles from ambient and cosmogenic radioactivity. A supernova near the Milky Way Center (10~kpc) will be easily detected inducing $\sim$60 measurable events, and the water tank will have a 3$\sigma$ sensitivity to supernovae up to 22~kpc, seeing $\sim$10 events. This paper shows how, even without dedicated optimization, modern dark matter experiments will also be able to play their part in the multi-messenger effort to detect the next galactic core-collapse supernova.

}

\begin{document}
\maketitle
\flushbottom
\section{Introduction}\label{Sec:Introduction}

Triggered by its own gravitational collapse, the death of a massive star is marked by the emission of a violent shock-wave (core-collapse supernova) and the birth of new compact objects, such as neutron stars or black holes. These supernovae are some of the most intense sources of neutrinos in our universe, as over a timescale of a few seconds, $\sim$99$\%$ of the stellar gravitational binding energy is converted into neutrinos with energies on the order of a few tens of MeV~\cite{mirizzi2016supernova}. SN1987A was the first supernova from which neutrinos were observed, but only about two dozen events were detected from this astrophysical phenomenon~\cite{arnett1989supernova}. The rate of galactic core-collapse supernovae has been predicted to be 1.63 $\pm$ 0.46 per century, but this rate is not uniform throughout the galaxy~\cite{rozwadowska2021rate}. One of the stars closest to earth that is likely to undergo a core-collapse supernova is Betelgeuse, which is located at 200~pc~\cite{kamland2024combinedpresupernovaalertkamland}. 

Observation of neutrinos from the next galactic core-collapse supernova will provide critical information about the neutrino driven explosion mechanism, the properties of neutrinos in ultra-dense environments, verify current hydro-dynamical modeling of stars, possibly provide evidence of new physics, and more~\cite{mirizzi2016supernova,raffelt2011supernova,raffelt2010physics}. As such, in the subsequent decades since the last supernova, many dedicated supernova neutrino observatories have been established~\cite{kuzminov2012baksan,zuber2015halo,salathe2012novel,aiello2021km3net,abe2016real,wang2022supernova} or are planned~\cite{abi2021supernova,juno2022juno,agostini2020pacific,pattavina2021res,abe2021supernova} with sensitivities varying over the local galactic group (kpc - Mpc). 

The current generation of supernova observatories primarily consists of large (a few m$^3$ to km$^3$), monolithic target detectors, and are mostly sensitive to $\nu_e$ and $\bar{\nu}_e$. Currently, the largest detectors sensitive to supernova neutrinos are Super-Kamiokande~\cite{SKSN} and IceCube~\cite{ICSN}. Both predominantly expect to detect $\bar{\nu}_e$ through inverse beta decay and some $\nu_e$ through charged current interactions. However, the detection of all other flavors is important for understanding the stellar mechanics of supernovae. In 2017, Coherent Elastic Neutrino Scattering (\CEvNS) was observed for the first time by the COHERENT experiment~\cite{akimov2017observation}. This interaction channel is sensitive to all flavors, has a high interaction cross-section (relative to other neutrino channels), and future experiments, such as RES-NOVA~\cite{beeman2023characterization,res2022radiopurity}, will seek to exploit this channel in the search for supernova neutrinos. In addition, direct detection dark matter experiments have also begun evaluating this channel for their sensitivity to galactic core-collapse supernovae~\cite{lang2016supernova,raj2020neutrinos,agnes2021sensitivity,ko2023sensitivities}.

The COSINUS (Cryogenic Observatory for SIgnatures seen in Next-generation Underground Searches) experiment~\cite{angloher2020cosinus,angloher2016cosinus} will operate sodium iodide (NaI) crystals as cryogenic scintillating calorimeters, with the primary purpose of cross-checking the DAMA/LIBRA experiment~\cite{bernabei2004dark,bernabei2008dama,bernabei2015final,bernabei2022further}. COSINUS is located at the Laboratori Nazionali del Gran Sasso (LNGS) under 1400~m of rock overburden (3600~m.w.e) and will feature a dual-channel read-out system of both heat and scintillation light. The dual-channel read-out system will allow for event-by-event discrimination between e$^-$/$\gamma$ interactions and nuclear recoils~\cite{angloher2023particle,COSINUS2023kqd}. The experiment is designed to minimize the amount of background nuclear recoils. This is achieved by choosing a site under a large overburden, developing low-radioactive crystals, ensuring that materials close to the detectors are as radiopure as possible, and passive~\cite{angloher2022simulation} and active shielding~\cite{angloher2024water}. This effort is designed to give a background rate of less than one nuclear recoil event per kg per year. COSINUS can search for \CEvNS\ of supernova neutrinos in the nuclear recoil spectrum of the NaI crystals, and since the active shielding is a water Cherenkov detector other interaction channels such as inverse beta decay can be explored.

This paper will investigate the neutrino interaction channels offered by the COSINUS experiment and its sensitivity to the next core-collapse galactic supernova. Section~\ref{Sec:Experimental_Setup} will detail the experimental setup as well as the different phases of the experiment. Section~\ref{Sec:Detection_Channels} will explore the neutrino detection channels available in both the NaI crystals and the water Cherenkov detector, while Section~\ref{Sec:Sensitivity} will discuss the galactic sensitivity of these channels to the next supernova. Finally, Section~\ref{Sec:Conclusion} will summarize the results presented herein.

\section{Experimental Setup}\label{Sec:Experimental_Setup}

The COSINUS experimental facility is built to house a dry dilution refrigerator with 48 superconducting quantum interference devices (SQUIDs) for amplification and read-out of the sensors. Inside the dilution refrigerator are copper boxes that contain the NaI detector modules. During the operation of the cryostat, the NaI detector modules will be operated at temperatures on the order of 10 mK. Besides reducing thermal noise, this allows for the use of transition edge sensors (TESs) made from tungsten films, to measure the phonon signal. A COSINUS NaI detector module consists of two parts: the NaI target equipped with a TES and a silicon beaker. Due to NaI's low melting point and softness, depositing a TES directly onto the absorber cannot be done. Instead, a TES is deposited on a sapphire wafer and is linked to a gold pad placed on the crystal with a thin gold wire. This setup is called the ``remoTES''~\cite{angloher2023first}. A TES is also deposited on the silicon beaker to measure the scintillation light of particle interactions. The design of the beaker allows for maximized light absorption. The combination of light and phonon channel allows for event-by-event particle discrimination, as well as setting
the absolute energy scale, which would be impossible with just one channel.

For the first phase of the experiment (referred to as COSINUS-1$\pi$), one box with 8 modules will be installed. Each module will have one (2.1~cm)$^3$ NaI crystal, for a total mass of 272~g of target material. This configuration will be used for the first year of data taking. Then in the second phase (COSINUS-2$\pi$), two more boxes are planned to be added. This will bring the total to 24 modules with a total target material mass of 816~g. If more exposure is required, 117~g hexagonal crystals could be installed in each of the 24 modules bringing the total up to 2808~g (COSINUS-Max). 

\begin{figure}[ht]
    \centering  \begin{subfigure}[t]{0.80\textwidth}
    \centering \includegraphics[width=\linewidth]{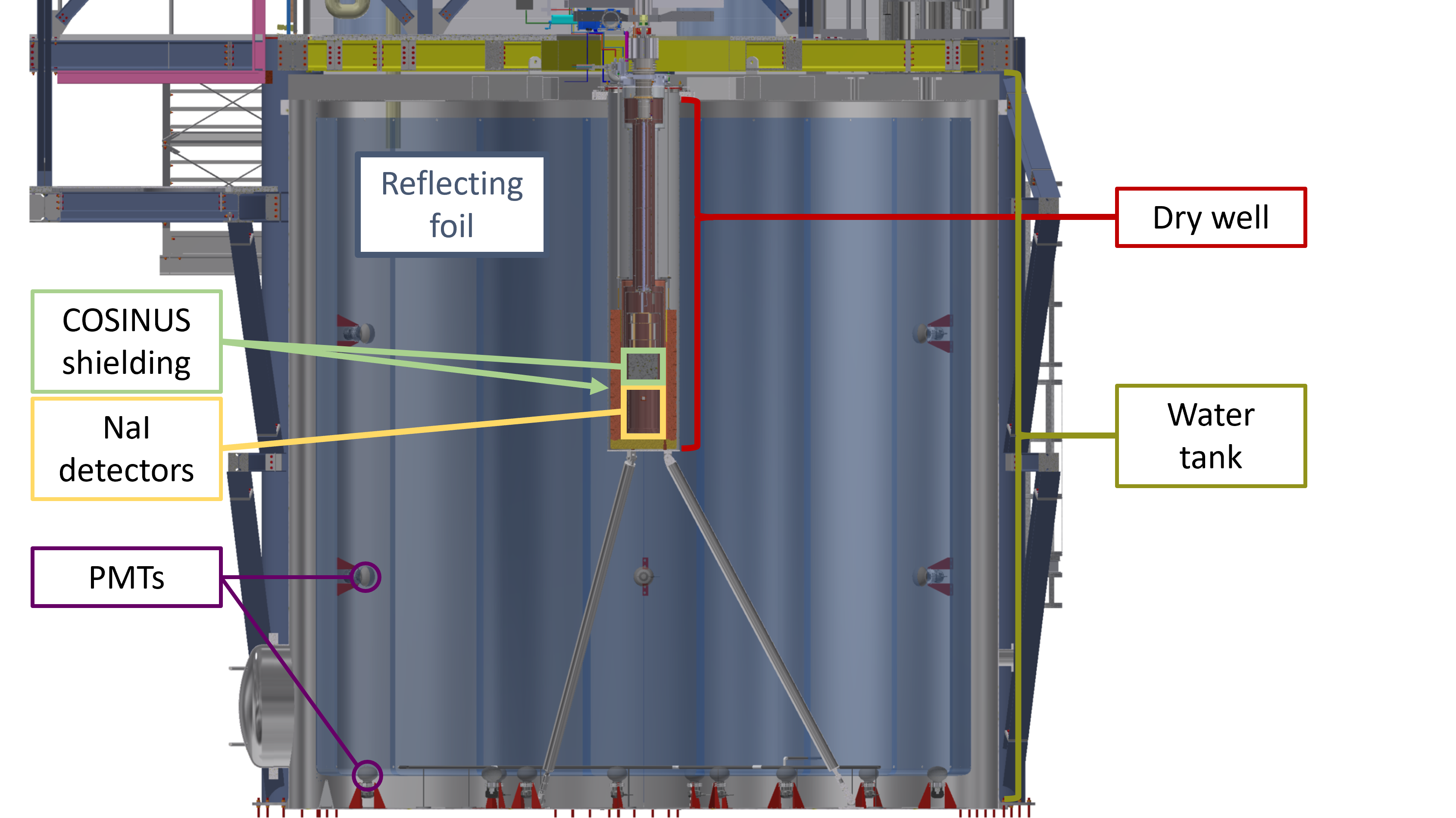}
    \caption{Labelled rendering of the COSINUS experimental setup pointing out regions described in Section~\ref{Sec:Experimental_Setup}. }
    \end{subfigure}
    \centering \begin{subfigure}[t]{0.75\textwidth}
    \centering \includegraphics[width=\textwidth]{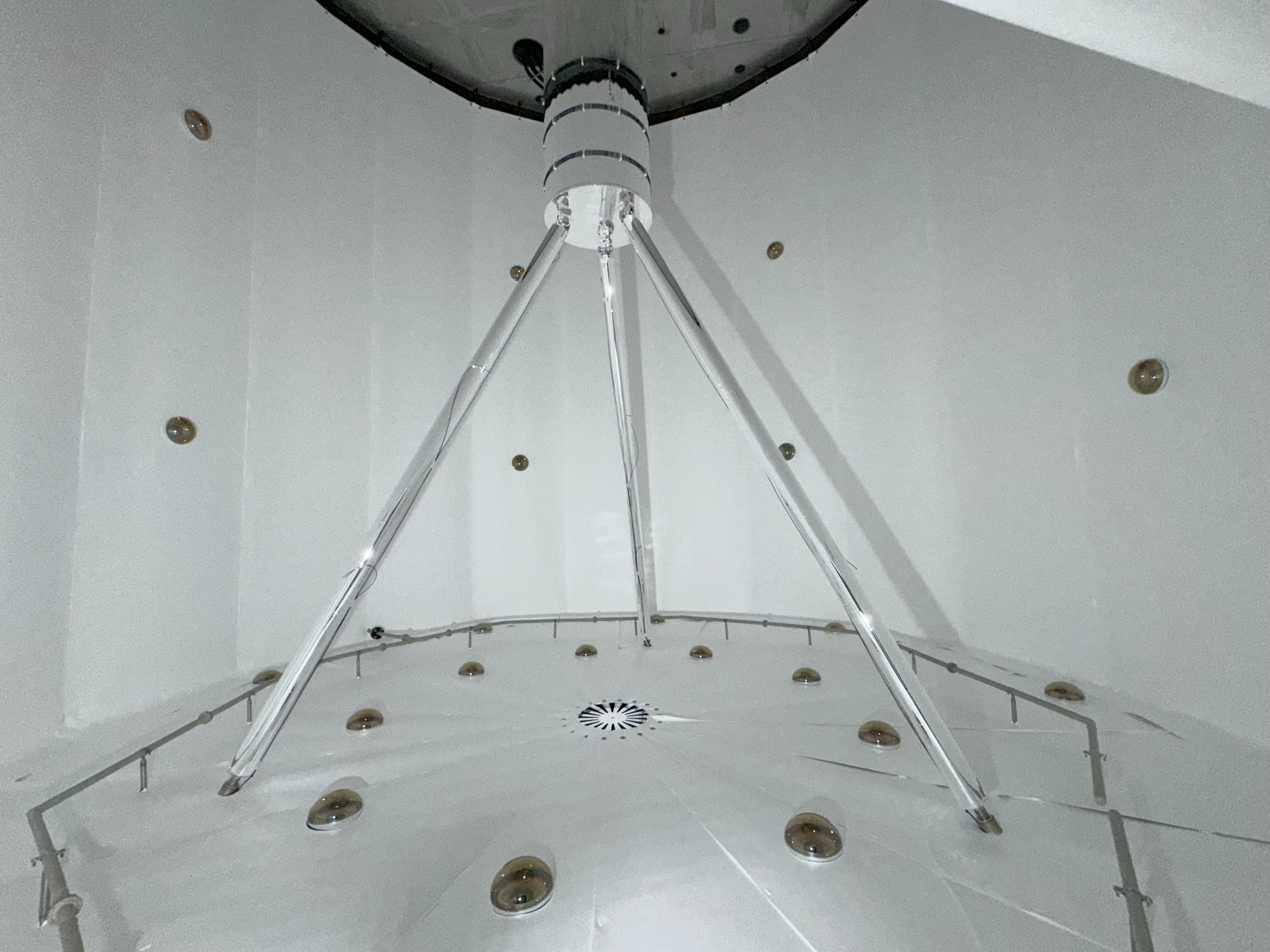}
    \caption{A photo of the inside of the muon veto tank without any water.}
    \end{subfigure}
    \caption{A schematic [a] and picture [b] of the COSINUS setup.}         \label{Fig:Supernova_Water_Tank.png}
\end{figure}

To minimize the background, additional shielding has been installed in the facility. This includes 30~cm of copper which is put inside the dilution refrigerator, above the detector boxes. During operation, there is an additional 8~cm of copper shielding surrounding the cryostat inside the drywell. The experimental setup including both copper shields can be seen in Fig.~\ref{Fig:Supernova_Water_Tank.png}.

A water tank is used as a passive and active shield for the experiment. During operation, the dilution fridge is lowered into a drywell so that the detector modules are surrounded by water without the outside of the cryostat getting wet. This drywell is 3.66~m deep and has a radius of 0.33~m. The water tank itself is 7~m tall with a diameter of 7~m and is filled with ultra-pure water. The active veto is accomplished with 30 photomultiplier tubes (PMTs) arranged inside the tank. The PMTs are 8-inch diameter R5912-30 from Hamamatsu~\cite{HamamatsuPhotonics}. Their arrangement can be seen on the floor and walls of the water tank in Fig.~\ref{Fig:Supernova_Water_Tank.png}. To account for the physical space of the PMTs and to reduce triggers from ambient gamma rays, a 30~cm dead layer was introduced near the walls of the water tank. Tyvek, a highly reflecting material, is hung with 30 cm gap from the wall
providing the dead layer and an improving light collection. The type of Tyvek used is 1082D, with two layers of Tyvek sandwiching a polyethylene layer. The reflectivity of this foil has been measured to be at least 95\%~\cite{adam2015juno}. The properties needed for the foil and the positioning of the PMTs were investigated using Monte Carlo simulations. The tank was optimized to be as efficient as possible as a muon veto, achieving muon veto efficiency of 99.63~$\pm$~0.16$\%$ and 44.4~$\pm$~5.6$\%$ for showers of secondary particles ($\sim$5$\%$ of all events). Giving a total efficiency of  97.0~$\pm$~0.3$\%$.~\cite{angloher2024water}. 

\subsection{Monte Carlo Modelling of the Water Tank}\label{SubSec:Monte_Carlo_Modelling}

The geometry of the setup has been implemented using a Geant4~\cite{GEANT4:2002zbu,Allison2006Geant,ALLISON2016186} based software toolkit called ImpCRESST~\cite{CRESST:2019oqe}, which was initially developed within the CRESST~\cite{abdelhameed2019first} dark matter search and later also within COSINUS~\cite{angloher2024water}. ImpCRESST utilizes Geant4 v10.2.3 and root v6-08-06~\cite{brun1997root}. In the simulation, the geometry is modeled as described in Sec.~\ref{Sec:Experimental_Setup}. A 30~cm dead layer is achieved with a reflective foil that was set to have 95\% reflectivity. Thirty PMTs are setup inside the tank with 18 being placed on the bottom and 12 along the wall. This is two more PMTs than what was studied in~\cite{angloher2024water} as the technical detail of the veto changed since that study. Along the bottom they are arranged into two concentric circles of 9 PMTs each, with the inner circle having a radius of 1.5~m and the outer having a radius of 3.2~m. Along the wall they are placed in two rows, evenly spaced, of 6, where the PMTs are equally distributed around the tank. The results of the ImpCRESST simulation consist of information such as particle tracking, energy deposition, particle type, interaction mechanism, and timing information. For this particular study, the propagation of the optical photons was also recorded. 

\clearpage
\section{Supernova Neutrino Interaction Channels in COSINUS}\label{Sec:Detection_Channels}

During a core collapse supernova $\sim$3$\times$10$^{53}$~ergs~\cite{Halim_2021} of neutrinos will be emitted within a short time window, approximately 10~s. This intensity over a small timescale makes these violent, astrophysical events excellent sources of neutrinos for terrestrial detectors. Interactions with these terrestrial detectors can take many forms (\CEvNS, inverse beta decay, charge current interactions etc.) and their sensitivity is highly target dependent.

\subsection{Supernova Neutrino Fluence Models}\label{subsec:Supernova_Neutrino_Fluence_Models}

\begin{figure*}[ht]
\begin{subfigure}[t]{0.50\textwidth}
\includegraphics[width=\textwidth]{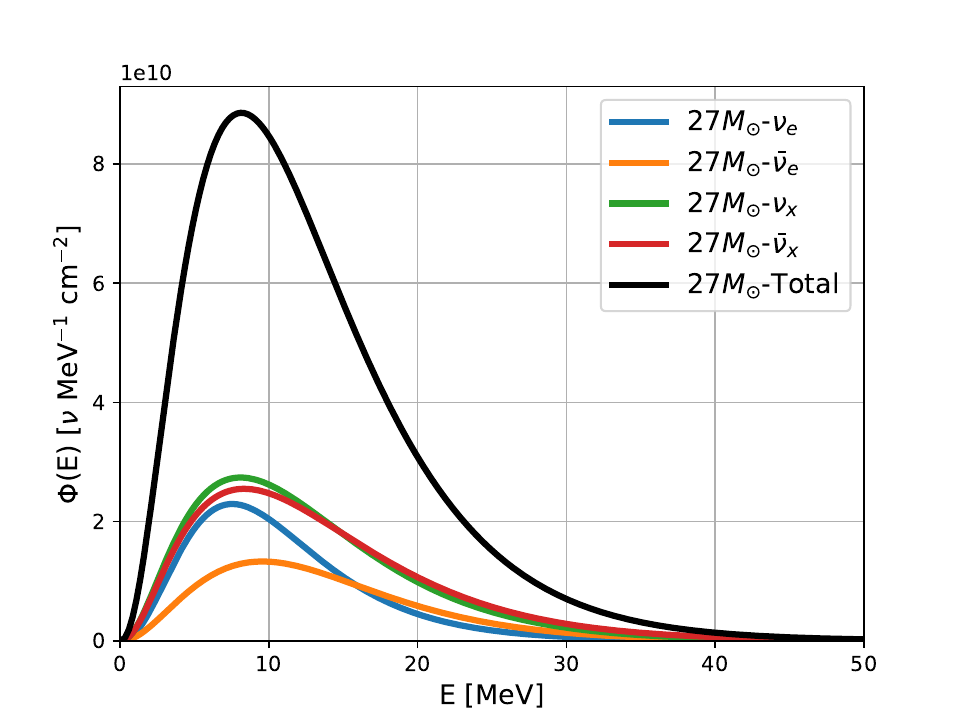} 
\caption{$27M_\odot$ simulation}
\end{subfigure}
\begin{subfigure}[t]{0.50\textwidth}
\includegraphics[width=\textwidth]{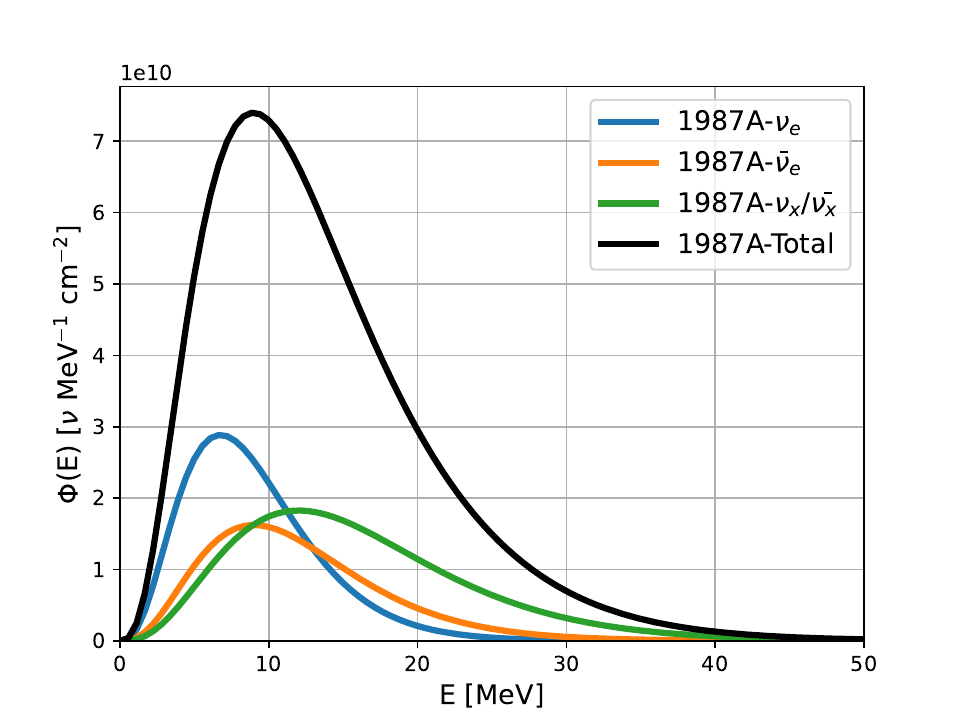}
\caption{1987A-like model}
\end{subfigure}
\caption{Neutrino flux expected from a core-collapse supernova at 10~kpc for two different models. Given that in the 1987A-like model the flux parameters are the same for both $\nu_x$ and $\bar{\nu}_x$ only their sum is shown.}\label{Fig:Supernova_Flux_10kpc}
\end{figure*}

The exact details of how the explosion takes place and how neutrinos are emitted is not fully understood and quite model dependent~\cite{mirizzi2016supernova}. Given this uncertainty this paper evaluates two different neutrino fluence models. The first adopts a quasi-thermal spectrum~\cite{Pagliaroli_2024,Keil_2003} to describe the differential neutrino fluence in energy: 
\begin{equation}\label{Eqn:Neutrino_Flux_1}
\Phi^d_i(E) = \frac{\epsilon_i}{4\pi d^2} \cdot \frac{E^\alpha e^{-E/T_i}}{T_i^{\alpha+2}\Gamma(\alpha+2)}, \quad  T_i = \langle E_{\nu_i} \rangle / (\alpha + 1)
\end{equation}
where $d$ corresponds to the supernova's distance from Earth and $E$ to the neutrino energy. Given that experiments can not distinguish between $\tau$ and $\mu$ supernova neutrinos, they are treated as having equal properties $\nu_\tau = \nu_\mu = \nu_x$. To have a comparison with already measured data, parameters corresponding to the best-fit for SN1987A events \cite{Halim_2021} were considered: $\langle E_{\nu_e} \rangle=9$~MeV, $\langle E_{\bar{\nu}_e} \rangle=12$~MeV, $\langle E_{\nu_x} \rangle=\langle E_{\bar{\nu}_x} \rangle=16$~MeV and $\alpha=3$. The total energy of the supernova shared equally between all neutrino flavors, is $\epsilon_i = 3/6$$\times$10$^{53}$~erg.
For the second model, an explosion with a $27M_\odot$ progenitor simulated by~\cite{Tamborra_2013_sim} was used. Due to these parameters varying in time within the simulation, only the time-averaged fluence was considered. These fluences can be seen in Fig. \ref{Fig:Supernova_Flux_10kpc}. 

\subsection{Neutrinos in the COSINUS Experiment}\label{subsec:Neutrinos_in_the_COSINUS_Experiment}

The COSINUS experiment can be broken down into two primary subsystems: the NaI cryogenic calorimeters and the water Cherenkov detector. Inside these subsystems the dominant interaction channels of impinging supernova neutrinos will be uniquely different.

\subsubsection{Neutrino Interaction Channels in NaI cryogenic calorimeters }\label{subsec:CEvNS}

For neutrinos interacting inside the NaI crystals, \CEvNS\ is the primary detection channel. Neutrinos will scatter off the nuclei causing recoils and depositing a few keV of energy into the target. The light output of nuclear recoils is quenched when compared to that of e$^-$/$\gamma$ events and can easily be distinguished by the COSINUS detector module setup~\cite{COSINUS2023kqd}. The target nuclear recoil background for the experiment will be less than one event per kg per year, so neutrino events corresponding to a galactic core collapse supernova are expected to coincide with a background-free environment. Additionally, \CEvNS\ is flavor independent and will be insensitive to neutrino oscillation. The differential recoil rate is given as

\begin{equation}\label{Eqn:Diff_Recoil}
    \frac{dR}{dE_r} = {\sum ^{Na,I}_t} {N_t}\int_{E^{\text{min}}_\nu} \Phi^d(E_{\nu}) \frac{d\sigma_t}{dE_r} dE_{\nu},
\end{equation}

where $N_t$ is the number of target nuclei and $\Phi^d(E_\nu)$ is the neutrino flux summed over all flavors defined by Eqn.~\ref{Eqn:Neutrino_Flux_1}.  The minimum energy required for a target nucleus to recoil is $E_\nu ^{\text{min}} = \sqrt{(E_r m_t /2)}$, where $E_r$ is the recoil energy of the nucleus, and $m_t$ is the target nucleus mass. The standard model differential cross-section~\cite{freedman1974coherent,abdullah2018coherent,abdullah2022coherent} for neutrino scattering requires the nuclear form factor to account for incoherency at higher neutrino energies. The Klein-Nystrand~\cite{sierra2019impact} form factor is used for this analysis,
\begin{equation}
    F_{KN} = \frac{2 j_1(qR_i)}{qR_i}\left(\frac{1}{1+q^2ak^2}\right),
\end{equation}
where $q = \sqrt{2m_nE_R}$ is the transferred momentum, the range of the Yukawa potential is set as $a_k$ = 0.7~fm~\cite{davies1976calculation} and $R_i = \sqrt{\frac{5}{3}R_{rms_i}^2-10a_k^2}$ where $R_{rms}$ is the root-mean square neutron (proton) radius and is taken as 2.92~fm (2.99~fm) for \Na~\cite{ohayon2022nuclear,angeli2013table} and 4.84~fm (4.75~fm) for \I~\cite{cadeddu2018average,angeli2013table}.

\begin{figure}[ht]
    \centering	\includegraphics[width=\textwidth]{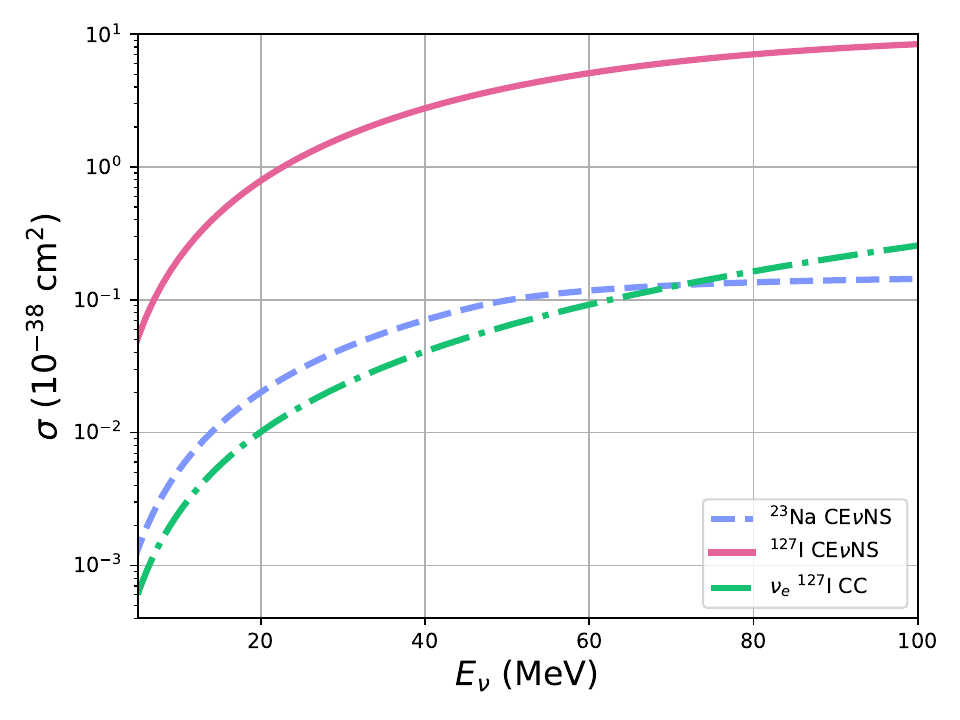}
	\caption{Cross-section comparison of selected neutrino interaction channels in NaI~\cite{freedman1974coherent,abdullah2018coherent,abdullah2022coherent,barbeau2023coherent,SNOWGLOBEs}. The neutrino-electron scattering cross-section in NaI is not shown but would be the same as in Fig.~\ref{Fig:Water_Tank_Cross_Sections.pdf}. See text for more details.}\label{Fig:NaI_Crosssection_Comparison.pdf}
\end{figure}

An additional channel of neutrino detection available in NaI crystals is charged current (CC)~\cite{COHERENTCCNAI} interactions with \I, defined as
\begin{equation}\label{Eqn:CCI_1}
  \nu_e + ^{127}\text{I} \rightarrow ^{127}\text{Xe} + e^-,
\end{equation}
and 
\begin{equation}\label{Eqn:CCI_2}
  \bar{\nu}_e + ^{127}\text{I} \rightarrow ^{127}\text{Te} + e^+.
\end{equation}
The neutrino energy threshold for lepton production will be $>$ 1~MeV, and the electron and positron produced will be visible in the e$^-$/$\gamma$ band. Given the large mass of the Te/Xe nuclei most of the neutrino energy becomes the lepton's kinetic energy ($\langle E_{\nu} \rangle \sim O(10)$~MeV). Detector efficiencies at such energies will be close to 100$\%$. Charged current interactions with $\tau$ or $\mu$ neutrinos are not possible as the energy of the supernova neutrinos will be below the required production threshold. Fig.~\ref{Fig:NaI_Crosssection_Comparison.pdf} compares the different interaction cross-section of neutrinos at supernova energies in NaI. The \CEvNS\ cross-sections are calculated directly from ~\cite{freedman1974coherent,abdullah2018coherent,abdullah2022coherent}\footnote{Where knowledge of physical parameters is required (e.g., $\theta_w$), the values are taken from~\cite{Workman:2022ynf} unless otherwise specified.} while the charged current cross-section are taken from SNOwGLoBES~\cite{barbeau2023coherent,SNOWGLOBEs}. It is clear from Fig.~\ref{Fig:NaI_Crosssection_Comparison.pdf} that within the NaI crystals the \CEvNS\ channel in \I\ will dominate the neutrino interactions by more than an order of magnitude. Other channels of neutrino interactions available for the NaI crystals (neutral current (NC) with \Na\ or \I\ or charged current with \Na) were not considered for this study. 

\subsubsection{Neutrino Interaction Channels in the Water Cherenkov Detector}\label{subsec:Water_Tank_Interactions}

When a charged particle travels through water at a speed faster than the speed of light in that medium Cherenkov radiation is produced~\cite{Jackson}. Whether through primary or secondary production neutrino interactions can produce charged particles of sufficient energy to create this radiation. Electrons and positrons must have a total energy above $\sim$0.77~MeV for Cherenkov radiation to be created. Table~\ref{tab:Neutrino_Water_Tank_Interactions} gives a summary of the possible neutrino-water tank interactions with any secondary reactions also being listed.
\begin{table}[ht]
    \centering
    \begin{tabular}{cccccc}
    \hline
    Process  & Interaction & Detectable & Cherenkov & Production \\
    &&& Threshold & Threshold \\\hline
    Proton Scattering & $\nu + p \rightarrow \nu' + p'$ &  No  & 1.4~GeV &  0 MeV \\\hline
    Electron Scattering     & $\nu + e^- \rightarrow \nu' + {e^-}' $  & Yes & 0.77~MeV & 0 MeV \\\hline
    Inverse Beta Decay & $\bar{\nu}_e+ p \rightarrow n + e^+$ & Yes & 0.77~MeV & 1.80~MeV\\
     & $n + p \rightarrow d + \gamma~(2.2$ MeV$)$ & No &  0.77~MeV & - \\\hline
    $\nu_e$ CC Oxygen       & $\nu_e + ^{16}$O$ \rightarrow ^{16}$F$ + e^-$ & Yes & 0.77~MeV & 15.9~MeV  \\\hline
    $\bar{\nu}_e$ CC Oxygen & $\bar{\nu}_e + ^{16}$O$ \rightarrow ^{16}$N$ + e^+$  & Yes & 0.77~MeV & 10.9~MeV \\\hline
    NC Oxygen  (\cite{tilley1993energy})    & $\nu + ^{16}$O$ \rightarrow \nu' + ^{16}$O$^*$ & No & - & -\\
     & $^{16}$O$^* \rightarrow ^{16}$O$ + \gamma$ & Yes & 0.77~MeV & 6.05~MeV \\\hline
    \end{tabular}
    \caption{Neutrino interactions inside the COSINUS water tank. When applicable any secondary interaction is also shown. When $\nu$ is present without a subscript this interaction is sensitive to all flavors of neutrinos. See the text for explanations of the reactions that are not detectable.}
    \label{tab:Neutrino_Water_Tank_Interactions}
\end{table}
The COSINUS water tank is sensitive to neutrino processes showcased in Table~\ref{tab:Neutrino_Water_Tank_Interactions}. Several other processes not listed in Table~\ref{tab:Neutrino_Water_Tank_Interactions} were considered, but the water tank is insensitive to these channels due to threshold or efficiency constraints. Firstly, proton scattering will not be visible in the water Cherenkov detectors due to its high threshold requirement for the production of Cherenkov radiation (1.4~GeV). Secondly, the subsequent 2.2~MeV gamma produced from the neutron capture is also not detectable due to the low efficiency of the COSINUS water tank at those energies, see Sec.~\ref{subsubsec:Detector_Efficiency}. The recoil of the scattered oxygen atoms in neutral current reactions will not create Cherenkov radiation and is invisible to our detector. Finally, there are many available excited states of $^{16}$O during the neutrino neutral current scattering interaction. The first excited state is at 6.05~MeV, corresponding to the minimum required neutrino energy for this process. Most of the higher level excited states will decay through the production of an alpha particle~\cite{tilley1993energy} and similar to the proton, this particle will be below the threshold energy of Cherenkov production. The exception will be the first five states which will de-excite through the production of $\gamma$-rays. The energy of the $\gamma$-rays will be between 6.05 - 7.11~MeV, so for the analysis in this study an average energy of 6.58~MeV will be assumed. 

\begin{figure}[ht]
    \centering	\includegraphics[width=\textwidth]{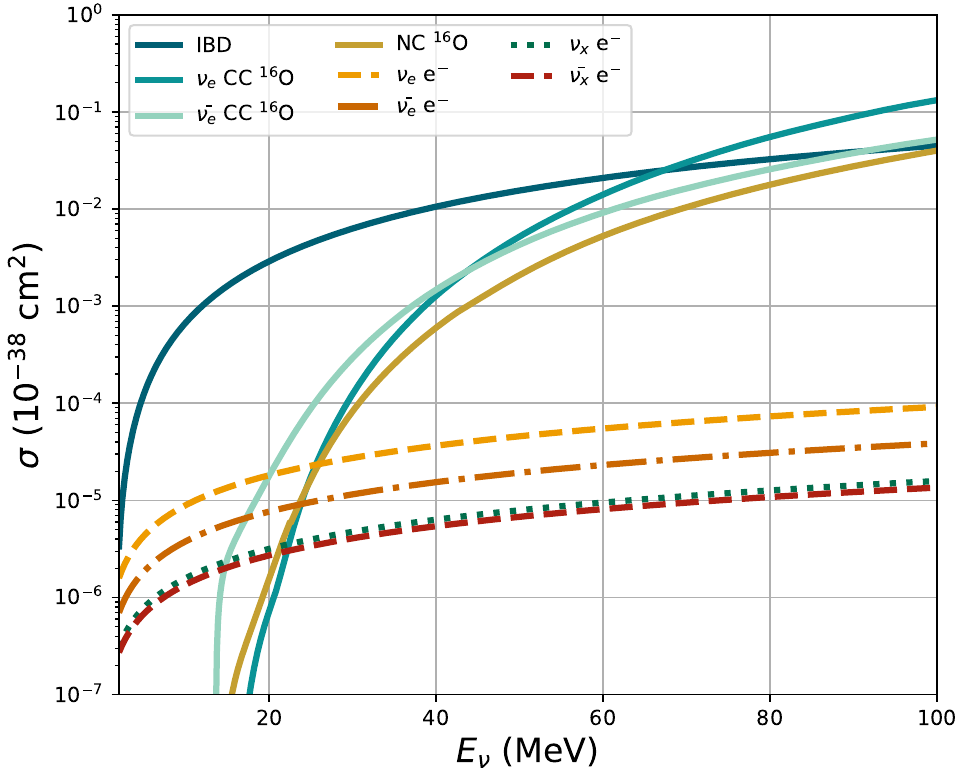}
	\caption{Cross-sections of the different neutrino interactions in the water tank. }\label{Fig:Water_Tank_Cross_Sections.pdf}
\end{figure}

For most of the detectable processes listed in Table~\ref{tab:Neutrino_Water_Tank_Interactions} the number of detected supernova events can be given by~\cite{strumia2003precise}

\begin{equation}\label{Eqn:Num_Expected_Supernova_Events}
    N_x = T_{x} \int_{E_{thresh_x}}\Phi^d_{\nu}(E_\nu) \sigma_{x}(E_\nu) \epsilon_x(E_\nu) dE_{\nu},
\end{equation}
where $T_x$ is the number of targets for the interaction $x$, $\Phi^d_{\nu}$ is the supernova neutrino flux defined in Sec.~\ref{subsec:Supernova_Neutrino_Fluence_Models}, $\sigma_{x}$ is the interaction cross-section, $\epsilon_{x}$ is the detection efficiency and $E_{thresh_x}$ is the threshold energy defined in Table~\ref{tab:Neutrino_Water_Tank_Interactions}. For the energy of supernova neutrinos the cross-sections of the interactions of Table~\ref{tab:Neutrino_Water_Tank_Interactions} are shown in Fig.~\ref{Fig:Water_Tank_Cross_Sections.pdf}, with the data taken from SNOwGLoBES~\cite{barbeau2023coherent,SNOWGLOBEs,becksnowglobes}. To estimate the event rate of electron-neutrino scattering using the cross section~\cite{giunti2007fundamentals}, an integral over both the neutrino energies and outgoing electron energies is required, similar to the procedure for calculating \CEvNS\ event rates. Electron scattering will contribute only a small percentage of the total number of neutrino interactions and will only be relevant for very close supernovae. Accounting for the neutrino oscillations for all flavor dependent cross sections is done following the process described by~\cite{dighe2000identifying,kato2017neutrino,seadrow2018neutrino}, which considers vacuum oscillations and the Mikheyev-Smirnov-Wolfenstein effect in the stellar envelope but not Earth-matter effects. Of note is the dependence of the survival probabilities on the mass hierarchy. It is clear that for the average expected energy of supernova neutrinos (10 - 20~MeV) inverse beta decay will be the primary channel of neutrino interaction. However, detector efficiency can increase or decrease the prominence of certain interaction channels, which is discussed in the following section.

\subsubsection{Water Cherenkov Detector Efficiency}\label{subsubsec:Detector_Efficiency}

\begin{figure*}[ht]
\begin{subfigure}[t]{0.32\textwidth}
\includegraphics[width=\textwidth]{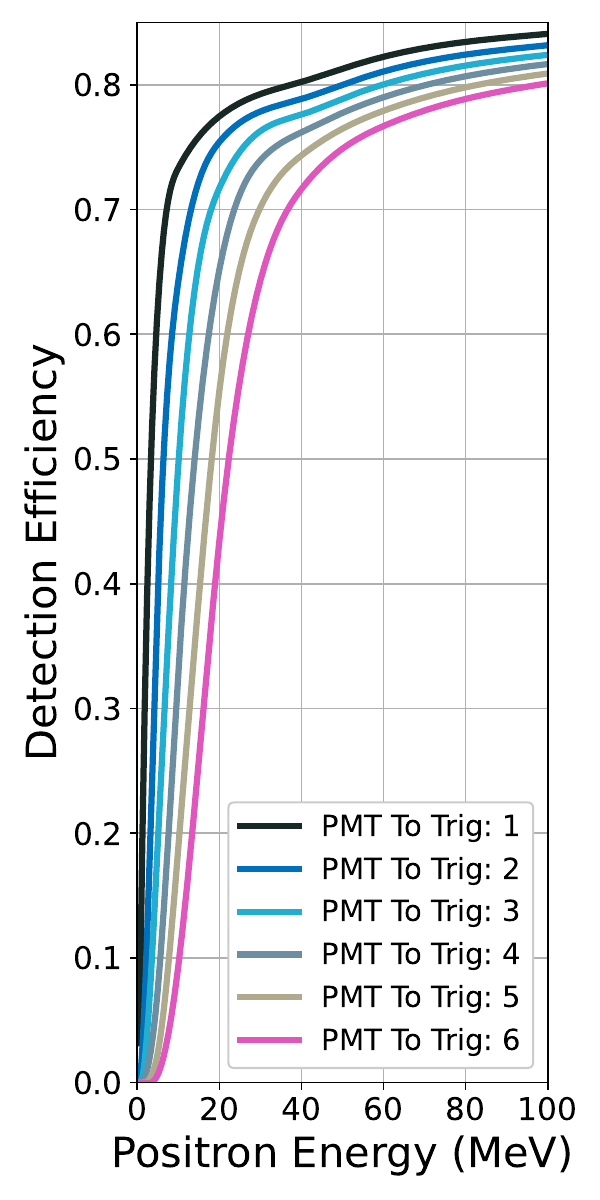} 
\caption{}
\end{subfigure}
\begin{subfigure}[t]{0.32\textwidth}
\includegraphics[width=\textwidth]{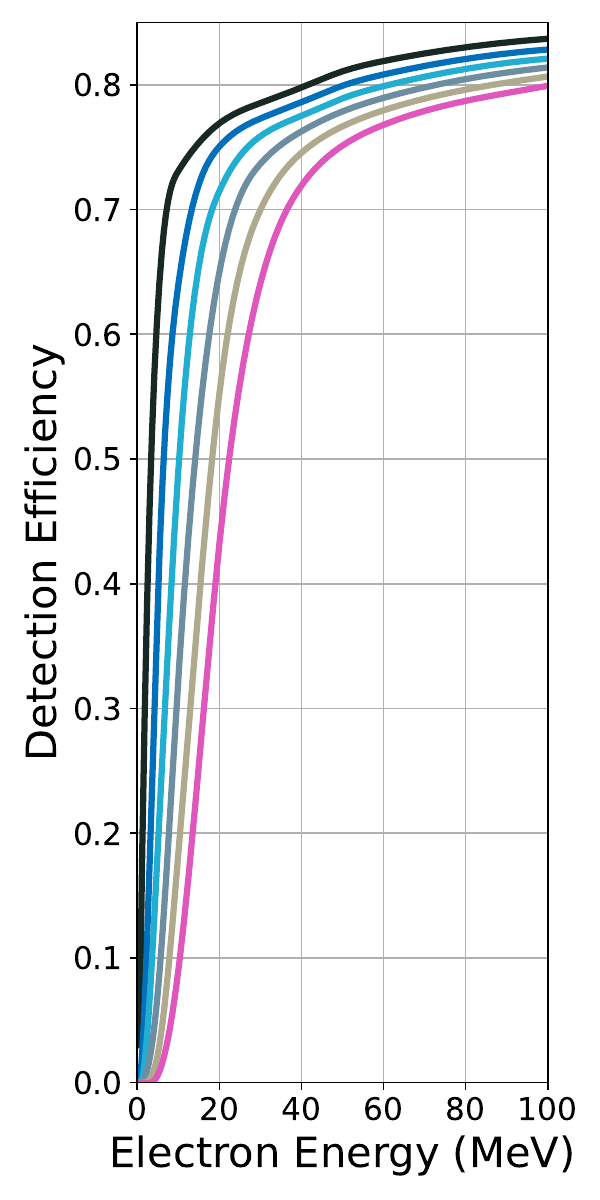}
\caption{}
\end{subfigure}
\begin{subfigure}[t]{0.32\textwidth}
\includegraphics[width=\textwidth]{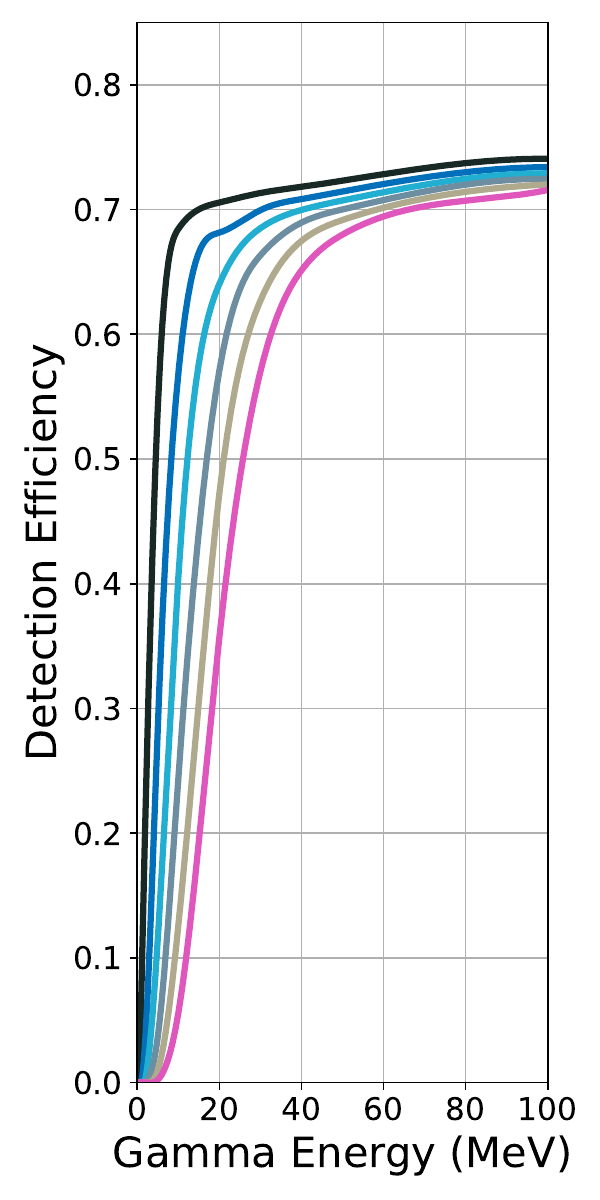}
\caption{}
\end{subfigure}
\caption{COSINUS water Cherenkov detection efficiency for (a) positrons (b) electrons and (c) $\gamma$ particles and different number of PMTs required to be in coincidence.}\label{Fig:Water_Tank_Detection_Efficiency.pdf}
\end{figure*}

Electrons, positrons and $\gamma$-rays are the detectable products of neutrino interactions within the water tank. As described in Sec.~\ref{SubSec:Monte_Carlo_Modelling}, the COSINUS water tank was modeled in Geant4 to evaluate the detection efficiency of these particles over a typical supernova energy range (1 - 100~MeV). The simulation was performed by choosing a monoenergetic energy (0.5, 1, 5, 10, 20, 30, 40, 50, 60, 80 or 100~MeV) and particle type (positron, electron or gamma) and uniformly distributing it inside the water tank volume. The particle momentum was isotropic, and the amount of light that arrived in the PMTs was recorded. Different trigger conditions (i.e. PMT threshold, PMTs in coincidence, etc.) could then be evaluated to determine if the particle was \textit{detected}. The efficiency is determined as the number of detected events over the total number of events simulated. This process was then repeated over all the above mentioned energies and particle types. For energies in-between the simulated energy values the efficiency is interpolated between the simulated data points. Fig.~\ref{Fig:Water_Tank_Detection_Efficiency.pdf} shows the energy dependent efficiency for positrons, electrons and $\gamma$-rays. For this analysis a single photo-electron threshold was used, as this will be the operating threshold of the COSINUS PMTs~\cite{angloher2024water}, and the quantum efficiency of the PMTs was selected to be 30$\%$, corresponding to the specifications of the manufacturer at the Cherenkov wavelength.  Fig.~\ref{Fig:Water_Tank_Detection_Efficiency.pdf} also shows how the efficiency varies depending on the number of PMTs that are required to trigger for an event to be classified as \textit{detected}. Based on concerns of high-background triggers the COSINUS experiment will require 4 - 6 PMTs to trigger during run-time conditions and a coincidence window of at least 500~ns. The efficiency for detecting positrons and electrons was found to be practically identical, while the $\gamma$-ray efficiency was about 10$\%$ less. Specifically, the efficiency for detecting the 2.2~MeV $\gamma$-ray from a neutron capture was found to be 0.071$\%$, making this channel unviable for this experiment. 

\section{Sensitivity to Core-Collapse Supernova Neutrinos
}\label{Sec:Sensitivity}
 \begin{figure}[ht]
    \centering
	\includegraphics[width=\linewidth]{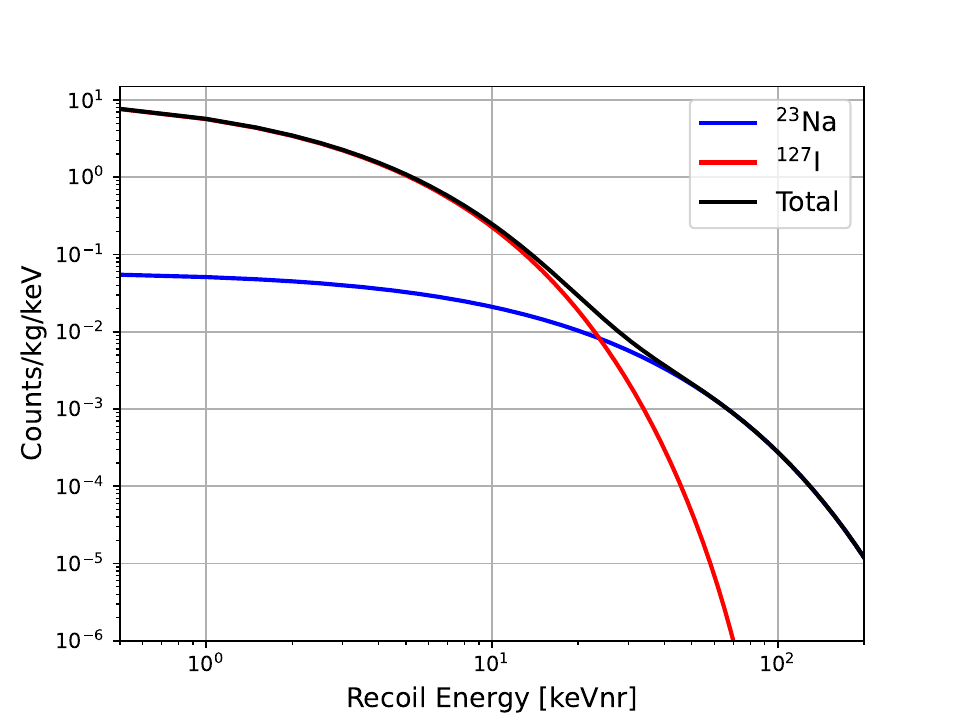}
    \caption{Nuclear recoil energy spectrum for NaI from \CEvNS\ from a 1987A-like  core-collapse supernova at a 400~pc distance.}\label{Fig:Nuclear_Recoil_Spectrum_NaI.pdf}
\end{figure}

Through the combination of a large, optically instrumented water tank and modular cryogenic NaI calorimeters, the COSINUS experiment will have sensitivity to a multitude of neutrino interaction channels. These include: \CEvNS, inverse beta decay, neutral and charged current interactions, and electron scattering. The following sections explore how the discovery significance of these channels will change as a function of distance to the supernova, and discuss how event pile-up can limit the sensitivity to near-supernovae.

\subsection{Detection in NaI Cryogenic Calorimeters}\label{subsec:NaI_Sensitivity}

The COSINUS detector modules, described in Sec.~\ref{Sec:Experimental_Setup}, will have the ability to distinguish nuclear recoils from other e$^-$/$\gamma$ interactions. With the target neutron background goal of less than one event per kg per year and the short ($\sim$10~s) length of supernovae, this will allow for zero background events and thus, direct identification of \CEvNS\ interactions inside the detector module. A trigger efficiency of 100$\%$ in the phonon channel of the NaI module is assumed for this analysis~\cite{COSINUS2023kqd}. By integrating over the neutrino energy in Eqn.~\ref{Eqn:Diff_Recoil}, Fig.~\ref{Fig:Nuclear_Recoil_Spectrum_NaI.pdf} shows the expected \CEvNS\ recoil energy spectra for a 400~pc supernova. Due to its higher mass, the \I\ isotopes will be the dominant target for \CEvNS\ interactions, while the \Na\ isotopes will be capable of recoils at higher energies. The target threshold for the COSINUS detectors is 1~keVnr, and by integrating the recoil energy spectrum above the threshold the total number of expected \CEvNS\ events can be determined. For a total detector mass of 1~kg and the recoil energy spectrum shown in Fig.~\ref{Fig:Nuclear_Recoil_Spectrum_NaI.pdf}, this gives 10.6~$\pm$~3.2 expected events. With zero expected background events during the timescale of a supernova the significance ($\sigma = \frac{n_s}{\sqrt{n_s + n_b}}$~\cite{bityukov2002uncertainties,abdullin1999search}) can be characterized as 3.2, or colloquially at a level indicating evidence of a supernova.

\begin{figure*}[ht]

\centering

\begin{subfigure}[t]{0.8\textwidth}
\centering	
\includegraphics[width=1.0\textwidth]{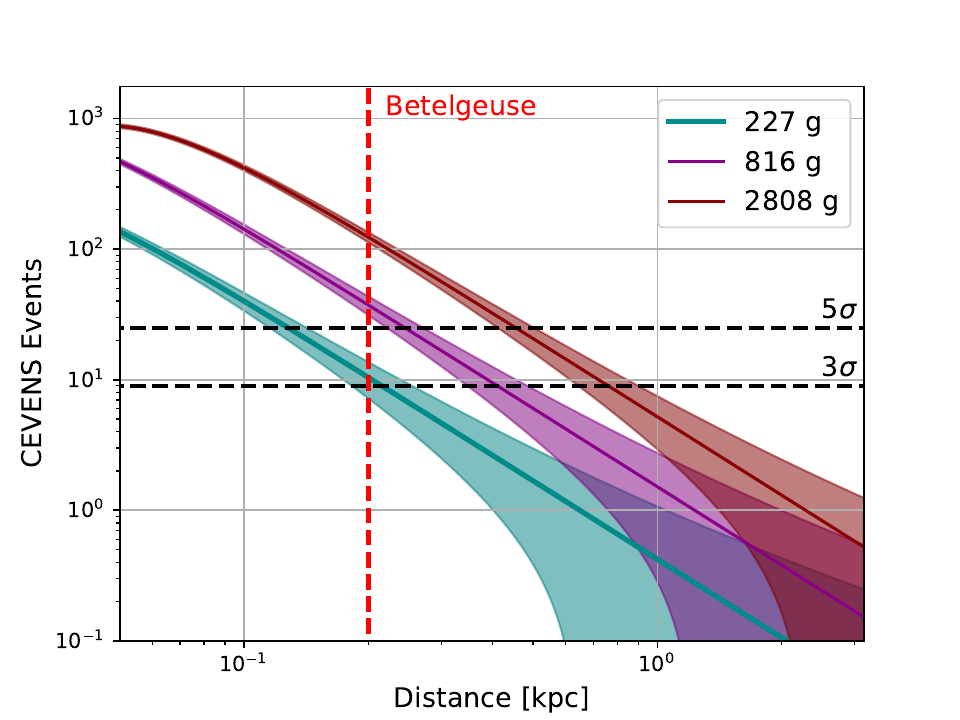}
\caption{$27M_\odot$ supernova
}
\end{subfigure}
\begin{subfigure}[t]{0.8\textwidth}
\centering	
\includegraphics[width=1.0\textwidth]{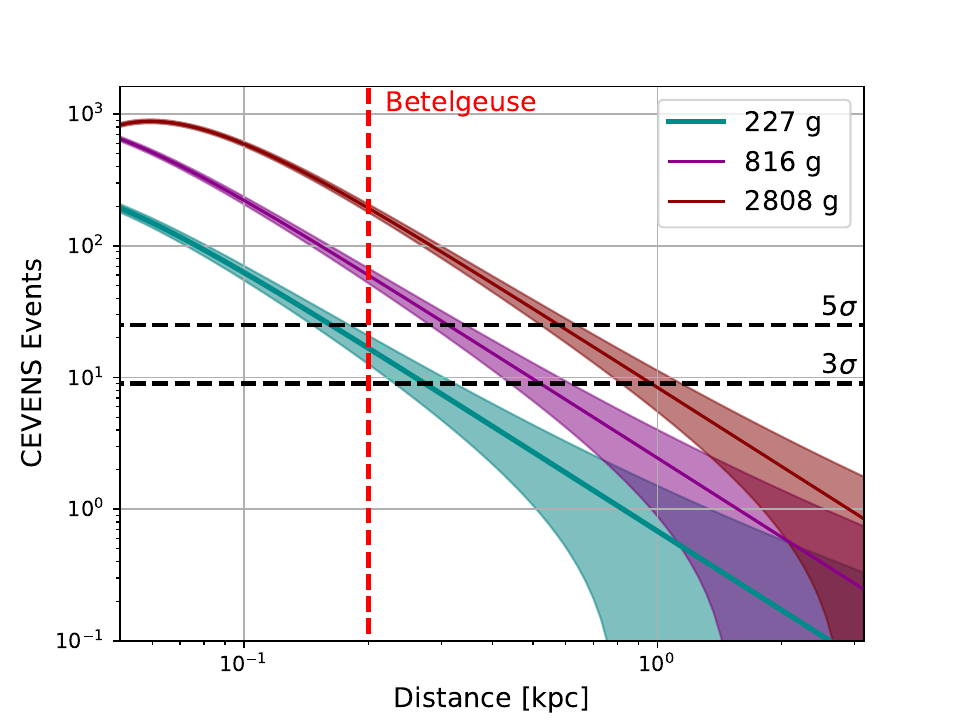} 
\caption{1987A-like supernova}
\end{subfigure}

\caption{Number of \CEvNS\ events from perspective near-supernovae in the NaI calorimeters, along with the statistical error at one standard deviation level. A 1~keV threshold was assumed and three different target masses were considered. The 3 and 5 sensitivity discovery potential is drawn as a horizontal band and the distance of the Betelgeuse star, a potential supernova candidate, is shown. Pile-up events that would arrive in a singular module during a typical pulse length have also been subtracted.}\label{Fig:CEvNS_Events_vs_Distance}
\end{figure*}

Fig.~\ref{Fig:CEvNS_Events_vs_Distance} shows the expected number of \CEvNS\ events, along with the statistical error, for an array of supernova distances for both considered models. Three different target mass phases of the COSINUS experiment: COSINUS-1$\pi$ (272~g), COSINUS-2$\pi$ (816~g), and COSINUS-Max (2808~g) are shown. The 3 and 5 significance band, denoting when \textit{evidence} or \textit{discovery} of a supernova can be claimed, are also overlaid. 

For a 1987A-like supernova the different target mass phases will be able to claim \textit{evidence} for a supernova at 0.300$^{+0.06}_{-0.05}$~kpc, 0.60$^{+0.08}_{-0.09}$~kpc, and 1.06$^{+0.19}_{-0.16}$~kpc, respectively. Signal pile-up inside the detector module will reduce the total number of events visible during a supernova and is taken into account in Fig.~\ref{Fig:CEvNS_Events_vs_Distance}. The probability of having a signal pile-up over a specified time window $\Delta T$ with a signal rate $R_{S}$ is given by 
\begin{equation}\label{Eqn:Pileup_Prob}
    p(R_S,\Delta T) = \int^{\Delta T}_{0} R_S e^{-R_S T} dT = 1-e^{-R_S \Delta T}.
\end{equation}
Assuming an average duration for a supernova of 10~s, that \CEvNS\ interactions are uniformly distributed in the number of modules for each target mass, and that the time window around a remoTES cryogenic event is $\sim$100~ms~\cite{angloher2023first} the probability of a signal event arriving within the same window as another event can be determined. In Fig.~\ref{Fig:CEvNS_Events_vs_Distance} the pileup events are removed from the total expected events, resulting in the bending of the \CEvNS\ event distribution seen at very near distances. Ultimately, due to the modular nature of the COSINUS NaI, the effect of pileup is only noticeable at supernova distance less then 0.1~kpc, corresponding to 6 (8\%), 24 (10\%) and 270 (30\%) events lost due to pileup for respectively  227~g, 816~g and  2808~g for a 1987A-like supernova.


The expected number of CC\I\ events was also investigated using the cross-section shown in Fig.~\ref{Fig:NaI_Crosssection_Comparison.pdf} and Eqn.~\ref{Eqn:Num_Expected_Supernova_Events}. Unfortunately, the number of events will not be significant at any relevant supernova distances. For example, for the COSINUS-2$\pi$ phase, at the Betelgeuse distance of 0.197~kpc, only $\sim$2 CC\I\ events are expected to be observed. However, knowledge of the CC\I\ cross-section is fairly limited with the COHERENT experiment just recently publishing the first measurement of the energy-inclusive cross-section~\cite{an2023measurement}. Therefore, in the event of a supernova, the e$^-$/$\gamma$ band should still be analyzed in search of this interaction channel. 

Although the distance to which the NaI crystals are sensitive is small compared to the size of the local galactic group (100~kpc), these modular detectors are excellent for detecting nearby, core-collapse supernovae, such as the candidates discussed in~\cite{mukhopadhyay2020presupernova}. This is due to their sensitivity to all flavors of neutrinos, lack of  backgrounds in the nuclear recoil band, and low-pileup threshold because of their modular nature.

\subsection{Detection in Water Cherenkov Detector}\label{subsec:Water_Tank_Sensitivity}

The sheer size of the water tank will allow for a supernova sensitivity to much larger distances than the crystals which reside inside. The number of expected events in the water tank can be determined by applying Eqn.~\ref{Eqn:Num_Expected_Supernova_Events}, the cross-sections in Fig.~\ref{Fig:Water_Tank_Cross_Sections.pdf}, the efficiencies in Fig.~\ref{Fig:Water_Tank_Detection_Efficiency.pdf} and the effect of neutrino oscillation to the available interactions in Table~\ref{tab:Neutrino_Water_Tank_Interactions}. The exception is the electron recoil interactions. For a supernova at 10~kpc, Table~\ref{tab:Number_of_Expected_Events_10kpc} gives the number of expected events per interaction type, for normal and inverse ordering scenarios and for a PMT triggering requirement of 4 and 6-fold. Oxygen NC interactions are given as upper limits due to the small number of expected events. The error given is purely statistical, as it was found that systematic errors of the efficiency and required measured parameters were negligible. As expected, it is clear from this table that inverse beta decay events will be the dominant contribution to the total number of neutrino events.

\begin{table}[ht]

    \centering
    \begin{tabular}{ccccc}
    \hline
    Process  & Ordering & PMT-fold & 1987A Events & $27M_\odot$ Events \\\hline
    Electron  & Normal  & 4 & 2 $\pm$ 1  & 2 $\pm$ 1\\
     Scattering& Normal  & 6 & 1 $\pm$ 1 & 1 $\pm$ 1\\
     & Inverse & 4 & 2 $\pm$ 1 & 2 $\pm$ 1\\
     & Inverse & 6 &  1 $\pm$ 1 & 1 $\pm$ 1\\\hline
    Inverse Beta   & Normal  & 4 & 34 $\pm$ 6 & 41 $\pm$ 6 \\
    Decay & Normal  & 6 & 22 $\pm$ 5 & 27 $\pm$ 5\\
    & Inverse & 4 & 63 $\pm$ 8 & 62 $\pm$ 8\\
    & Inverse & 6 & 43 $\pm$ 7 & 42 $\pm$ 6\\\hline
    $\nu_e$ CC    & Normal  & 4 & 1 $\pm$ 1 & 1 $\pm$ 1\\
    Oxygen & Normal  & 6 & 2 $\pm$ 1 & 1 $\pm$ 1\\
    & Inverse & 4 & 1 $\pm$ 1 & 1 $\pm$ 1 \\
    & Inverse & 6 & 1 $\pm$ 1 & 1 $\pm$ 1\\\hline
    $\bar{\nu}_e$ CC  & Normal & 4 & 1 $\pm$ 1 & 1 $\pm$ 1\\
     Oxygen& Normal  & 6 & 1 $\pm$ 1 & 1 $\pm$ 1 \\
    & Inverse & 4 & 3 $\pm$ 2 & 2 $\pm$ 1\\
    & Inverse & 6 & 3 $\pm$ 2 & 1 $\pm$ 1\\\hline
    NC Oxygen  & Normal & 4  & $<$ 1 & $<$ 1\\
    & Normal  & 6 & $<$ 1 & $<$ 1 \\
    & Inverse & 4 & $<$ 1 & $<$ 1\\
    & Inverse & 6 & $<$ 1 & $<$ 1\\\hline
    \end{tabular}
    \caption{Number of expected neutrino interactions in the COSINUS water tank for two supernova models at 10~kpc. Both normal and inverse ordering are considered. For the efficiencies in Fig.~\ref{Fig:Water_Tank_Detection_Efficiency.pdf}, the 4 and 6 PMT triggering conditions are shown as this would represent the typical running state of the COSINUS experiment.}
    \label{tab:Number_of_Expected_Events_10kpc}
\end{table}

\begin{figure*}[ht]
\begin{subfigure}[t]{0.50\textwidth}
\includegraphics[width=\textwidth]{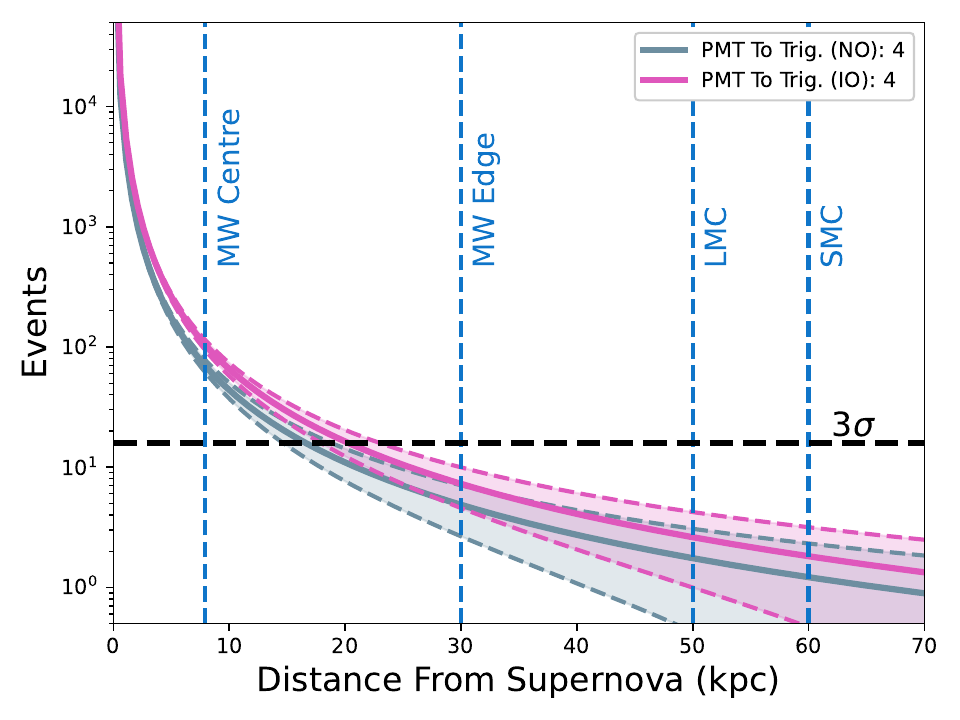} 
\caption{$27M_\odot$}
\end{subfigure}
\begin{subfigure}[t]{0.50\textwidth}
\includegraphics[width=\textwidth]{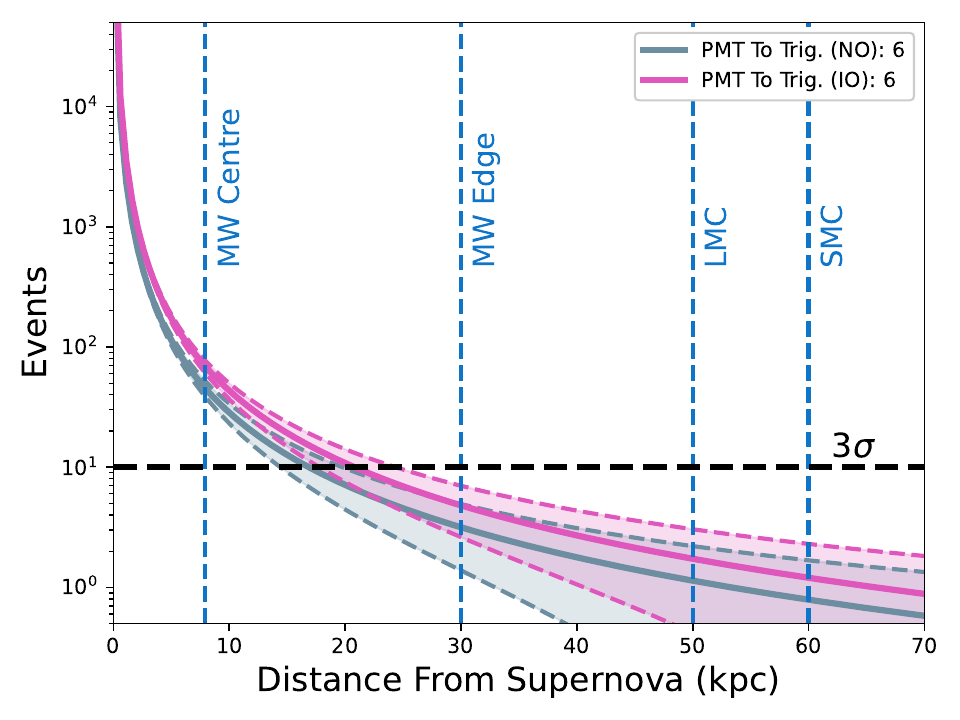}
\caption{$27M_\odot$}
\end{subfigure}
\begin{subfigure}[t]{0.5\textwidth}
\includegraphics[width=\textwidth]{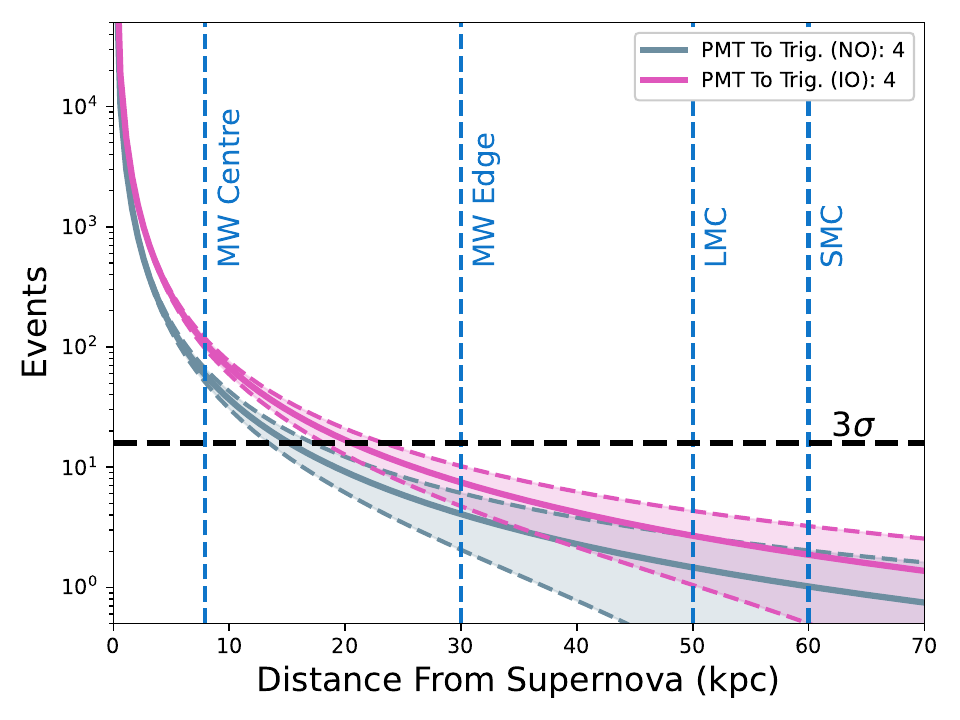} 
\caption{1987A-like}
\end{subfigure}
\begin{subfigure}[t]{0.5\textwidth}
\includegraphics[width=\textwidth]{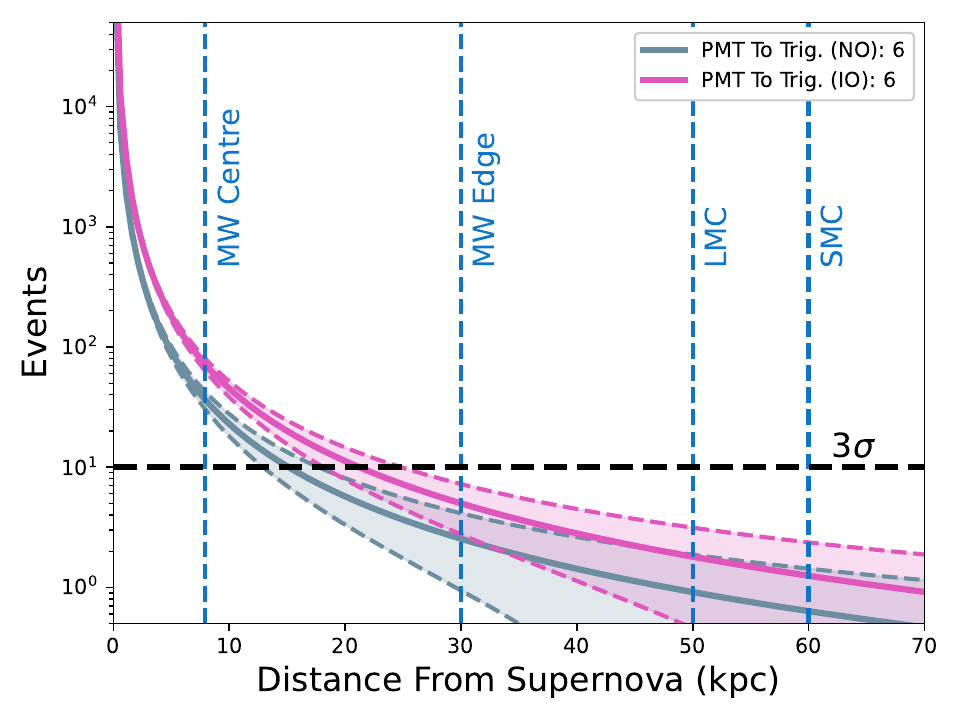}
\caption{1987A-like}
\end{subfigure}
\caption{Expected number of neutrino interactions in the COSINUS water Cherenkov detector. Upper graphs (a)(b) correspond to $27M_\odot$ data, and lower graphs (c)(d) correspond to 1987A-like data. Both the normal and inverted ordering case were considered for 4 (a)(c) and 6 (b)(d) PMT trigger threshold. Each plot shows the effect of the normal ordering (NO) and the inverse ordering (IO) of the neutrino hierarchy.} The 3$\sigma$ discovery potential is drawn as a horizontal line and the distance of the important galactic reference points are shown. Signal events that would arrive in the tank during a typical PMT pulse length have also been subtracted from the distribution.\label{Fig:Total_Water_Events_vs_Distance}
\end{figure*}

Fig.~\ref{Fig:Total_Water_Events_vs_Distance} shows the number of total expected events in the water tank as a function of supernova distance. Both the normal and inverse ordering are considered, as well as 4 and 6-fold PMT triggering conditions. The 3$\sigma$ significance denoting the evidence of a supernova is also overlaid. Similar to the analysis in Sec.~\ref{subsec:NaI_Sensitivity}, the pileup events have been removed from the total with an assumed trigger window of 1~$\mu$s, based on the tank characterization from~\cite{angloher2024water}. 
Unlike the NaI calorimeters the water tank will have a natural background rate, which has been characterized in~\cite{angloher2024water} and consists of muon events, ambient radiation and thermionic emissions inside the photocathodes of the PMTs. For a 4(6)-fold PMT triggering requirement the background rate was determined to be 1.05~Hz (0.09~Hz). 

The maximum distance at which the water tank could claim \textit{evidence} of a supernova as a function of supernova model and PMT triggering condition can be seen in Tab.~\ref{tab:Horizon vs Model}. In the most optimistic scenario the water tank will be sensitive up to $22^{+4}_{-3}$~kpc considering a 1987A-like supernova and inverted ordering. The horizon distance increases by up to $23\%$ when requiring 6 PMTs to be triggered as opposed to 4, this clearly shows how in that configuration the reduction of background is favored despite reducing the number of signal events. The number of PMTs that are required to trigger a signal does reduce the total event rate, however the large decrease in background rate allows the water tank to extend the horizon by a few kpc. This allows the water tank to be sensitive to supernovae just at the edge of the Milky Way galaxy.

\begin{table}[ht]

    \centering
    \begin{tabular}{c|cc|cc}
    \hline
    Model & \multicolumn{2}{|c|}{$27M_\odot$} & \multicolumn{2}{c}{1987A} \\
    \hline
    PMT-fold & 4 & 6 & 4 & 6 \\
    NO horizon (kpc) & $17^{+2}_{-2}$ & $18^{+3}_{-3}$ & $15^{+2}_{-2}$ & $16^{+3}_{-2}$ \\
    IO horizon (kpc) & $21^{+3}_{-2}$ & $22^{+4}_{-3}$ & $21^{+3}_{-2}$ & $22^{+4}_{-3}$
    
    \end{tabular}
    \caption{Sensitivity horizon for a $3\sigma$ detection for both supernova models, PMT operating modes and orderings.}
    \label{tab:Horizon vs Model}
\end{table}

\section{Conclusion}\label{Sec:Conclusion}
The COSINUS dark matter experiment has been evaluated for its sensitivity to the next galactic core-collapse supernova. Through modular, cryogenic, NaI calorimeters and a large water Cherenkov detector the COSINUS experiment can probe supernovae both near and far in the Milky Way. With \CEvNS\, the NaI crystals will be sensitive to all flavors of neutrinos from supernovae up to 1~kpc. Their modular nature minimizes the effects of pileup and the ability for event-by-event particle discrimination creates a background free environment during a supernova explosion. Supernova neutrino interactions in the water tank will be dominated by inverse beta decay events and for the assumptions considered in this study the COSINUS water tank could be sensitive to a supernova between 15 and 22~kpc away. Future work will involve integrating the COSINUS experiment into the Supernova Early Warning System~\cite{al2021snews} and evaluating the directional sensitivity of the water Cherenkov detector. 
Moreover, as discussed in \cite{Pagliaroli_2024}, COSINUS can be able to identify the time of black-hole formation in case of Failed CCSNe with an uncertainty of a few ms, providing a useful alert for the nearby gravitational wave interferometer Virgo to look for the GW counterpart.
Finally, the COSINUS experiment, with one large homogeneous detector containing modular cryogenic calorimeters in the center, is distinctly sensitive to almost all of the possible neutrino interactions in matter (\CEvNS, CC, NC, and electron scattering). Although COSINUS was not optimized for the detection of supernova neutrinos, the variety of interaction channels it is sensitive to and the possibility of coincidence analysis could represent a design philosophy to follow for future dedicated supernova neutrino experiments.




\bibliographystyle{unsrt}
\bibliography{biblio.bib}
\end{document}